\documentclass[12pt]{article}
\usepackage{amssymb,amsmath,epsfig,multicol}
\usepackage{graphicx}

\def\today{\number\day\space\ifcase\month\or
January\or February\or March\or April\or May\or June\or July\or August\or
September\or October\or November\or December\fi \space\number\year}

\makeatletter

\def\@begintheorem#1#2{\trivlist \item[\hskip \labelsep{\sc #1\ #2.}]\sl}
\def\@opargbegintheorem#1#2#3{\trivlist
\item[\hskip \labelsep{\sc #1\ #2\ (#3).}]\sl}
\def\@endtheorem{\endtrivlist}

\def\@sect#1#2#3#4#5#6[#7]#8{\ifnum #2>\c@secnumdepth
\let\@svsec\@empty\else
\refstepcounter{#1}\edef\@svsec{\csname the#1\endcsname.\hskip 0.5em}\fi
\@tempskipa #5\relax
\ifdim \@tempskipa>\z@
\begingroup #6\relax
\@hangfrom{\hskip #3\relax\@svsec}{\interlinepenalty \@M #8\par} \endgroup
\csname #1mark\endcsname{#7}\addcontentsline {toc}{#1}{\ifnum
#2>\c@secnumdepth \else
\protect\numberline{\csname the#1\endcsname}\fi #7}\else
\def\@svsechd{#6\hskip #3\relax\@svsec #8.\csname #1mark\endcsname
{#7}\addcontentsline
{toc}{#1}{\ifnum #2>\c@secnumdepth \else \protect\numberline{\csname
the#1\endcsname}\fi #7}}\fi
\@xsect{#5}}

\newcommand{\be}{\begin{equation}}
\newcommand{\bT}{\begin{array}}
\newcommand{\eT}{\end{array}}
\newcommand{\de}{\end{equation}}
\newcommand{\ee}{\end{equation}}

\def\text#1{{\quad \hbox{#1} \quad}}

\def\ds{\displaystyle}

\begin{document}
\title{A pseudosimilarity approach to a steady free convection flow }
\author{
 \textsc{Zakia Hammouch}\footnote {Email adress: zakia.hammouch@u-picardie.fr, Tel: +33 03 22 8 78 47} \\
{\footnotesize \it LAMFA, CNRS UMR 6140, Universit\'e de Picardie
Jules Verne,
}\\
{\footnotesize \it Facult\'e de Math\'ematiques et d'Informatique,
33, rue Saint-Leu 80039 Amiens, France}
\date{}}
\maketitle
\noindent \rule{17cm}{0.02cm} {\bf Abstract}\newline
 {\footnotesize \noindent 
In this communication we deal with the exact solutions called "pseudosimilarity" of a steady free convection problem studied by by Kumaran and Pop (2006). They showed that there is no similarity solution for the case of a wall temperature as $T_{w}(x)\sim x^{-\frac{1}{2}}$ (resp. a wall heat flux as $q_{w}(x)\sim x^{-\frac{3}{2}},$ and a dimensionless heat transfer coefficient $h_{w}(x)\sim x^{-1}$). We shall present some results about existence and asymptotic behaviour of new exact solutions of the resulting boundary value problem for each case.\\}
\noindent\rule{17cm}{0.02cm}

\section{Introduction}
\setcounter{equation}{0} \setcounter{theorem}{0}
\setcounter{lemma}{0} \setcounter{remark}{0}
\setcounter{corollary}{0}
Heat and mass transfer in saturated porous media occurs widely in natural phenomena and industrial applications, such as extrusion of polymers, continuous casting, colling of metallic plates,etc. More details about the existing literature on this subject can be found in the books (Neild and Bejan, 1999; Vafai, 2000; Pop and Ingham, 2001) and the references therein.\\
Recently Kumaran and Pop (2006) presented some original results about  a particular case of this phenomena. They considered a fluid (water) characterized by a relationship between the temperature $T$ and the density $\rho$
\begin{equation}\label{0}
\frac{\rho - \rho_{c} }{\rho_{c}} = -\gamma(T-T_c)^{2},
\end{equation}
where $\rho_c$ is the maximum density in the liquid phase, and $\gamma$ is the fluid thermal expansion coefficient of water at $4^{\circ}C$. We note that equation (\ref{0}) have been obtained by Goren (1966). This kind of problems have been the subject of many papers, see for instance (Soundalgekar, 1980; Black el al., 1984; Poulikakos, 1984). \\
In (Kumaran and Pop, 2006), the authors studied the free convection about an impermeable flat plate embedded in a porous medium filled with water near the vicinity of its density maximum associated with the temperature of $4^{\circ}C$ at atmospheric pressure with the condition $u(x,\infty)=0$ (uniform free stream velocity). They showed analytically and numerically, the existence of multiple similarity solutions to the governing boundary layer equations for a class of problems namely, wall temperature (VWT,\, $T_{w}(x)=x^{m}$), variable heat flux (VHF,\, $q_{w}(x)=x^{\frac{4m-1}{2}})$, or variable heat transfer coefficient (VHTC,\, $h_{w}(x)=x^{\frac{2m-1}{2}}$). They also gave a nonexistence result for ($m=-\frac{1}{2}$), this singular case will be the main  goal of the present investigation. We shall adopt a pseudosimilarity approach to construct exact solutions for the resulting boundary value problem and to study their precise asymptotic behaviour.\\ 
According to Pop and Ingham (2001),  the physical model is described  by the following equations
\begin{equation}\label{1}
\left\{\begin{array}{ll}
u_{x}+v_{y} = 0,\\
u = T^{2},\\
uT_{x}+vT_{y} = T_{yy},
\\
v(x,0) = v_{w}x^{p},\\
u(x,\infty) = T(x,\infty) = 0.\\ 
 \end{array} \right.
\end{equation}
Supplemented by one of the following conditions
\begin{equation}\label{2}
\left\{\begin{array}{ll}
T(x,0)=  x^{m} &\quad (VWT),\\
T_{y}(x,0)=  x^{\frac{4m-1}{2}} &\quad (VHF),\\
-\frac{1}{T}T_{y}(x,0)= x^{\frac{2m-1}{2}} &\quad (VHTC).\\
 \end{array} \right.
\end{equation}
Where $u$ and $v$ are the velocity components along $x$ and $y$ axes, respectively, $T$ is the fluid temperature and the exponent $m$ is a  real number.\\
In the usual way, we introduce  the stream function $\psi, $ (satisfying $u=\psi_{y}$ and $v=-\psi_{x}$). Then the above model can be expressed in a simpler form 
\begin{equation}\label{3}
\left\{\begin{array}{ll}
\psi_{y} = T^{2},\\
\psi_y T_x-\psi _xT_y = T_{yy},\\
\psi_{x}(x,0)= -v_w x^{p}, \, \lim_{y\rightarrow \infty}\psi _{y} (x,y)= 0. 
 \end{array} \right.
\end{equation}
Define the similarity transformations
\begin{equation}\label{4}
\psi(x,y)=x^{r}f(\eta), \quad T(x,y)=x^{m} \theta(\eta), \quad \eta=y x^{s},
\end{equation}
where $f$ is the dimensionless stream-function, $\theta$ is the dimensionless temperature, $\eta$ is the similarity variable, and the real numbers $r$ and $s$ satisfy the relations $r+s=1$, $r-s=2m$ and $p=\frac{2m-1}{2}$. Injecting (\ref{4}) in (\ref{3}) we get
\begin{equation}\label{5}
\left\{\begin{array}{ll}
f'=\theta^{2},\\
\theta''+\frac{1+2m}{2}f\theta '-mf'\theta =0,\\
f(0)=f_{w}, \quad \theta(\infty)=0,
 \end{array} \right.
\end{equation}
subject to one of the following conditions
\begin{equation}\label{6}
\left\{\begin{array}{ll}
\theta(0)=1 &\quad (a),
\\
\theta'(0)=-1 &\quad (b),
\\
\theta'(0)+\theta(0)=0 &\quad (c).
 \end{array} \right.
\end{equation}
Where $f_{w}=-\frac{2}{2m+1}v_w$ (the suction/injection parameter), and the prime denotes differentiation with respect to $\eta$.\\
The boundary value problem (\ref{5}) , for $m=-\frac{1}{2}$, under the condition (i), $i={a,b,c}$, will be the studied in detail in the next section.
\section{The pseudosimilarity solution}
\setcounter{equation}{0} \setcounter{theorem}{0}
\setcounter{lemma}{0} \setcounter{remark}{0}
\setcounter{corollary}{0}
Magyari et al. (2002)  have studied the well-known free convection boundary layer flow problem about a vertical flat plate with an inverse-linear temperature distribution $T_{w}(x)=T_{\infty}+\frac{\alpha}{x}$ where $\alpha>0$, under the condition $u(x,\infty)=0$. They have shown that the similarity solutions in the classical form are missing, and proved that in order to overcome this difficulty, the stream function $\psi$ has to be shifted by the term $\sigma \log(x),\, \sigma>0$ in order to get exact solutions. In this section we adopt the same concept to obtain exact solutions for problem (\ref{4})-(\ref{5}), under the condition (\ref{3}), when $m=-\frac{1}{2}$. First, we study the case of a Variable Wall Temperature.\\

\noindent$\bullet$ {\bf The (VWT) flow}\\

\noindent In this situation, the wall temperature is given by 
\begin{equation}\label{8}
T_{w}(x)=x^{-\frac{1}{2}}.
\end{equation}
 Following the work  (Magyari et al., 2002), we define the new stream-function
\begin{equation}\label{9}
\psi(x,y)=F(\eta,x).
\end{equation}
If $m=-\frac{1}{2}$ we deduce from the previous section that($r=0, s=1$) and then the pseudosimilarity transformations are defined by
\begin{equation}\label{10}
\Psi (x,y)=F(\eta,x), \quad T(x,y)=x^{-\frac{1}{2}} \Theta(\eta,x), \quad \eta =y x^{-\frac{1}{2}},
\end{equation}
where $F$ is the modified dimensionless stream-function, $\Theta$ is the dimensionless temperature and $\eta$ is the pseudo-similarity variable, substituting in (\ref{4}) we obtain
\begin{equation}\label{11}
\left\{\begin{array}{ll}
F_{\eta}=\Theta^{2},\\
\\
\Theta_{\eta \eta }+\frac{1}{2}F_{\eta}\Theta = x F_{\eta}\Theta_{x}-x F_{x}\Theta_{\eta}.
 \end{array} \right.
\end{equation}
As in (Ibrahim and Hassanien, 2001), the term $\Theta_{x}$ can be neglected and we may assume that $\theta=\Theta(\eta,x)$ (then the term $F_{x\eta}$ can also be neglected). By writing $F(\eta,x)=f(\eta)+h(x),$ equation (\ref{3}) yields
\begin{equation}\label{12}
\left\{\begin{array}{ll}
f'=\theta^{2},\\
\theta''+\frac{1}{2}f'\theta =-x h'\theta'.
 \end{array} \right.
\end{equation}
Hence there exists a real number $\sigma$ such that 
\begin{equation}\label{13}
\left\{\begin{array}{l}
f'=\theta^{2},\\
xh'=\sigma,\\
\theta''+\frac{1}{2}f'\theta+\sigma \theta'=0. 
 \end{array} \right.
\end{equation}
The integration of the second equation gives $h(x)=\sigma \log(x)+cst,$ and then the new stream-function is given by $\psi(x,y)=f(\eta)+\sigma \log(x)+cst$. Coming back to problem (\ref{3}), we see that $\sigma$ has a physical meaning since $\sigma=-v_{w}$ (the suction parameter).\\
In the remainder, we deal with the following boundary value problem
\begin{equation}\label{14}
\left\{\begin{array}{l}
f'=\theta^{2},\\
\theta''+\frac{1}{2}f'\theta+\sigma \theta'=0,\\ 
\theta(0)=1, \quad \theta(\infty)=0.
 \end{array} \right.
\end{equation}
Let us notice that the no-slip condition $f(0)$ is not required, in fact it can be any real number (see (Hammouch, 2006) for more details). On the other hand, the substitution of (\ref{14})$_{1}$ in (\ref{14})$_{2}$ gives an unforced Duffing equation (Panayotoukanos et al., 2002)
\begin{equation}\label{15}
\theta''+\frac{1}{2}\theta^{3}+\sigma \theta'= 0 \\
\end{equation}
with the conditions
\begin{equation}\label{16}
\theta(0)=1, \quad \theta(\infty)=0.
\end{equation}
whose study in the phase plane reveals the stability of the unique equilibrium point $(0,0)$ is related to the sign of $\sigma$. Fisrt of all we show that $\sigma$ has to be positive, for this sake we define the Lyapunov function for equation (\ref{15}) by
\begin{equation}\label{17}
V(\eta)=\theta'+\frac{1}{4}\theta^{4},
\end{equation}
which is positive and satisfies $V'(\eta)=-\sigma{\theta'}^{2}$. If we assume that $\sigma\leq 0$, $V$ is monotonic increasing  and then condition $\theta(\infty)=0$ could not be satisfied. Then,  the real number $\sigma$ is assumed to be  positive, it follows that the equilibrium point $(0,0)$ is
globally asymptotically stable and  all local solutions of (\ref{15})-(\ref{16}) remain bounded and tend to zero at infinity.\\
 However, we remark that we cannot exclude the existence of oscillating solutions for smallest values of $\sigma$ (see Figure. 1). 
\begin{center}
 \includegraphics[width=10cm]{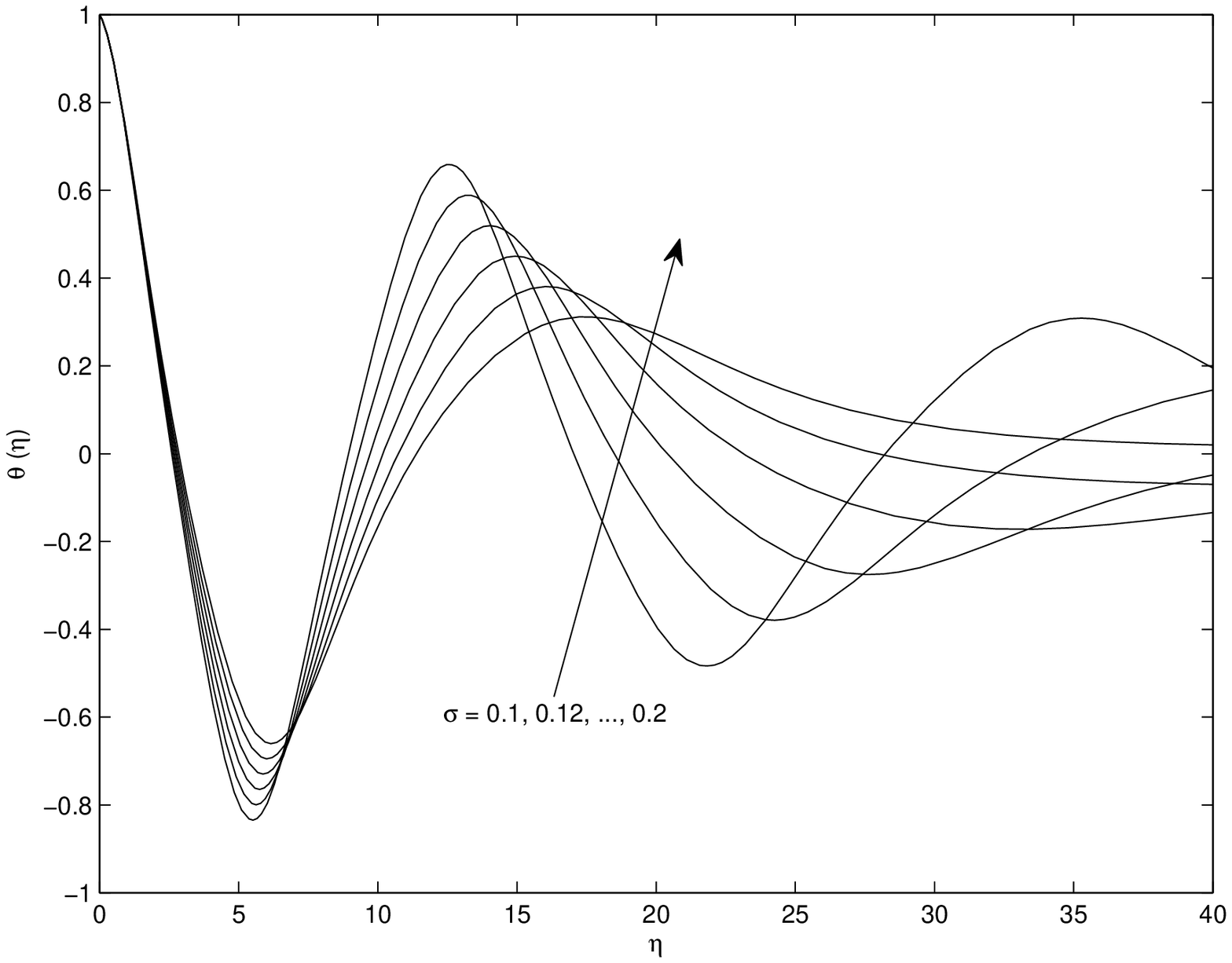}
 \end{center}
 \begin{center}
 {\footnotesize
 {Figure.$1$    Temperature Profiles  as a function of $\eta$ for small values of $\sigma$ and  $\theta'(0)=-0.1$, for the case (VWT).}}
 \end{center}
To establish the existence of monotonic  decreasing positive solutions to (\ref{15})-(\ref{16}), we shall adopt a Crocco variables approach (Nachman and Callegari, 1980; Ishimura and Ushijima, 2004). For the sake of simplicity, we assume that the function $\theta(\eta)$ (which can be regarded as an independent variable) is strictly monotonic $(\theta'<0$, for $\eta>0)$. Actually, solutions which are increasing on some interval $(0,\eta_0)$ may exist, although this case can be studied in an analogous way.\\
 Let $\eta(\theta)$ denote the inverse function of $\theta(\eta)$. Defining 
\begin{equation}\label{18}
s=\theta,\qquad \mbox{and} \qquad \phi(s)=-\theta'(\eta(s)),
\end{equation}
leads to
\begin{equation}\label{19}
\phi\phi^\prime = \sigma \phi - \frac{1}{2}s^{3}\quad 0 < s < 1,
\end{equation}
under the conditions
\begin{equation}\label{20}
\phi(0) = 0\qquad\mbox{and}\qquad \phi(\infty)=0.
\end{equation}
Integrating over $(0,s)$ leads to
\begin{equation}\label{21}
\phi(s) = \sigma s - \frac{1}{2}\int_0^s\frac{\tau{3}}{\phi(\tau)}d\tau,\quad s\in(0,1).
\end{equation}
To classify the pseudosimilarity solutions to (\ref{4})-(\ref{5}), we study the solutions of the more simpler integral equation 
(\ref{21}) with the conditions (\ref{20}). We notice that there is some analogy between problem (\ref{15})-(\ref{16}) and the following one:\\
Find the traveling-wave (TW) solutions $U(x,t) = \theta(x-\sigma t),$ with the condition $\theta(\infty)=0$, (where $\sigma$ is the wave speed), for a special case of the Fisher-KPP equation
\begin{equation}\label{22}
U_t = U_{xx}+ U^{3}.
\end{equation}
In the light of the work by Gilding and Kersner (2005) concerning this kind of problems (2.15), we deduce the existence of a threshold value $\sigma^{\star}$ such that existence and uniqueness of solution to (\ref{15})-(\ref{16}) is guaranteed if $\sigma=\sigma^{\star}$. While for $\sigma>\sigma^{\star}$ there are multiple solutions.\\
Now we pay attention to get  estimation for $\sigma^{\star}$, and to study the precise asymptotic behavior of $\theta$ as $\eta \rightarrow \infty$. Multiplying equation (\ref{19}) by $\displaystyle \frac{2}{s}$ leads to
\[\left( \frac{\phi^{2}}{s}\right) '+\frac{(\phi-\sigma s)^{2}}{s^{2}}+{s^{2}}=\sigma^{2},\] then
\[\left(\frac{\phi^{2}}{s}\right)'+\frac{1}{s}\leq \sigma^{2}.\]
A simple integration over $(0,1)$ gives
\begin{equation}\label{23}
{\phi(1)}^{2}+\frac{1}{3}\leq \sigma^{2},
\end{equation}
consequently
\begin{equation}\label{24}
\sigma^{\star} \geq \frac{1}{\sqrt{3}}.
\end{equation}
Now we proceed to determine the large $\eta$-behavior of these solutions. For this sake, we will  exploit an idea of Br\'ezis et al. (1986). First we show that for large $\eta$ we have the following limits
\begin{equation}\label{25}
\quad \lim_{\eta\rightarrow \infty}\frac{\theta'}{\theta}= -\sigma \,\,\,(\star) \qquad \mbox{or}\qquad  \lim_{\eta\rightarrow \infty}\frac{\theta'}{\theta}= 0 \,\,\,(\star\star).
\end{equation} 
Actually, setting $\chi(\eta)=\frac{\theta'(\eta)}{\theta(\eta)}$ gives
\begin{equation}\label{26}
\chi'+\sigma \chi+\chi^{2}+\frac{\theta^{2}}{2}=0,
\end{equation} 
for all $\eta\in(0,\infty)$. Because that $\phi(s)\leq \sigma s$ we get that the function $\chi$ is negative and bounded ($-\sigma \leq \chi(\eta) < 0)$, then for $\eta$ large $\chi$ has a finite limit, say $l$. This limit satisfies $l^{2}+l\sigma = 0$ which leads to (\ref{25}). \\
Assume now that $(\star)$ holds, then we have for all $s\in(0,1)$
\begin{equation}\label{27}
\frac{1}{s^{2}}[\frac{\phi(s)}{s}-\sigma]=-\frac{{\int_{0}}^{s}(r^{3}/\phi(r))dr}{2s^{3}}.
\end{equation}
By the  L'H\^{o}pital rule, we get
\begin{equation}\label{28}
\lim_{s\rightarrow 0}\frac{1}{s^{2}}[\frac{\phi(s)}{s}-\sigma]=-\frac{1}{6\sigma},
\end{equation}
then
\begin{equation}\label{29}
\phi(s)\sim \sigma s + \frac{s^{3}}{6\sigma}\qquad \mbox{as}\quad s\rightarrow 0
\end{equation}
Next, we assume that ($\star\star$) holds, from equation (\ref{19}) we get immediately
\begin{equation}\label{30}
\phi(s)\sim \frac{s^{3}}{2\sigma} \qquad \mbox{as}\quad s\rightarrow 0.
\end{equation}
Coming back to problem (\ref{15})-(\ref{16}) we conclude that\\
- For $\sigma = \sigma^{\star}$ : \qquad $\theta \sim  e^{-\frac{\sigma^{\star}}{\eta}}$ \qquad (exponential decay).\\
- For $\sigma > \sigma^{\star}$ : \qquad $\theta \sim \sqrt{\frac{\sigma}{\eta}}$  \qquad\,\, (algebraic decay).\\

\noindent In the remainder we shall obtain  some estimates  for $ \phi(1) $. We stress that the real number $ \phi(1)=-\theta'(0)$ plays the role of the shooting parameter for problem  (\ref{15})-(\ref{16}). We note also that equation (\ref{23}) gives a lower estimation for $\theta'(0)$, actually we have $\displaystyle \theta'(0)=-\phi(1)\geq-\sqrt{\sigma^{2}-\frac{1}{3}} $. \\
To obtain an  upper estimation  for  $ \theta'(0) $   we  look  for  a positive solution to (\ref{15})-(\ref{16}) such that $ \zeta := \theta '(0) > 0.$ In the above analysis, we have said that for any $  \zeta\not=0 $  the (unique) local solution to (\ref{15})-(\ref{16}) with the initial condition  $ \theta (0) = 1,$ is global and goes to zero at infinity.  Since  $ \zeta  $ is positive there exists a positive number  $ \eta_\zeta$ such that  $ \theta  $ is monotonic increasing on $(0,\eta_\zeta )$ and $ {\theta}'(\eta_\zeta) = 0.$ Setting $\lambda = \theta(\eta_{\zeta})$ and $\theta_{\zeta}=\theta(\eta+\eta_{\zeta})$
for all $ \eta  \geq 0.$  The shifted function $ \theta_\zeta $ is a solution to (\ref{15})-(\ref{16}) with the initial condition
\[  \theta _\zeta(0)= \lambda > 1,\quad  {\theta'}_{\zeta}(0) = 0.\]
Using the  transformations
\[ \eta \to \lambda^{2}\eta,\qquad\mbox{and}\qquad \theta_{\zeta} \to \frac{\theta_{\zeta}}{\lambda}.\]
Clearly the function $\theta_{\zeta}$ is a solution to the following problem
\begin{equation}\label{31}
\left\{\begin{array}{ll}
\theta''_{\zeta} + \sigma_{\lambda} \theta'_{\zeta} + \lambda{\frac{{\theta_{\zeta}}^{3}}{2}}=0,\\
\\
\theta_{\eta}(0)=1, \qquad \theta'_{\zeta}(0)=0,\quad  \theta_{\zeta}(\infty)=0,
\end{array} \right.
\end{equation} 
where $ \sigma_\lambda = \lambda^{-2}\sigma.$  \\
Thanks  to the above analysis, problem (\ref{31}) has positive solution if and only if
\begin{equation}\label{32}
\sigma > {\frac{\lambda}{\sqrt{3}}}. 
\end{equation}
On the other hand, integrating (\ref{31}) over $ (0,\eta_{\zeta})$ gives
\[ \ds \sigma \lambda + \frac{1}{2}\int_0^{\eta_\zeta}{\theta_{\zeta}^{3}}(\eta)d\eta = \zeta+\sigma.\] From which we deduce that 
\( \sigma \lambda < \zeta+\sigma.\)
Hence inequality (\ref{32}) is satisfied if
\begin{equation}\label{33}
\zeta  \leq \sigma\left( \sigma\sqrt{3}-1\right).
\end{equation}
In conclusion, under the (VWT) condition, problem (\ref{15})-(\ref{16})-(a) admits multiple solutions (flows). Every solution is uniquely parametrized by the skin-friction parameter $\theta'(0)\in
\left[-\sqrt{\sigma^{2}-\frac{1}{3}}, \sigma\left(\sigma\sqrt{3}-1\right) \right] $.
The above results have been illustrated numerically in Figure.2. 
\begin{center}
 \includegraphics[width=11cm]{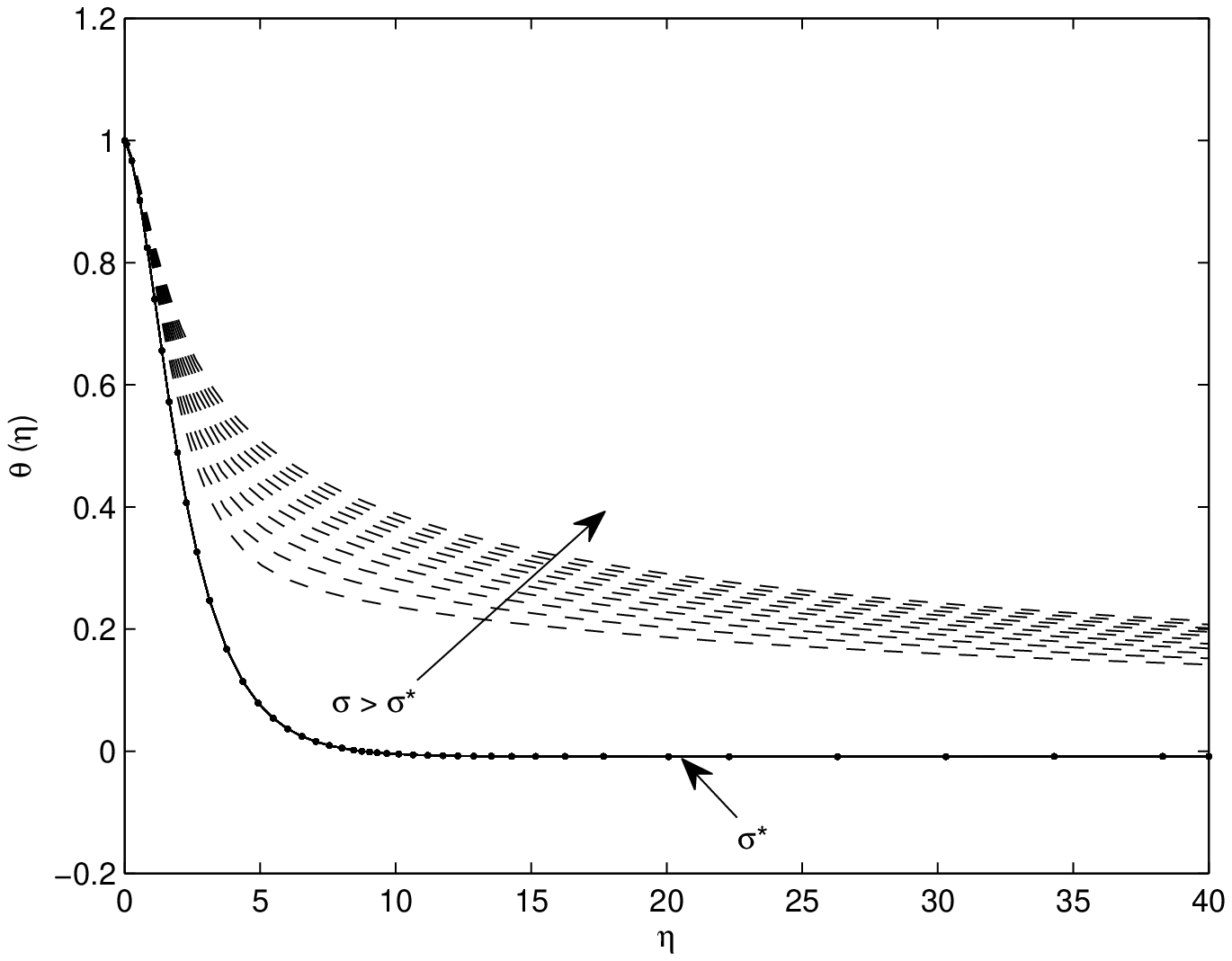}
 \end{center}
 \begin{center}
 {\footnotesize
 {Figure.$2$   Temperature Profiles  as a function of $\eta$ for various values of $\sigma$ and related values of $\theta'(0)$, for the case (VWT).}}
 \end{center}
$\bullet$ {\bf The (VHF) flow}\\

\noindent This flow  is characterized by the wall heat flow $q_w(x)=x^{-\frac{1}{2}}$ and the wall temperature $T_w(x)=0$. In this case, we study problem (\ref{15}),(\ref{16}) supplemented by condition $(b)$.\\  First we note that if $\theta$ is a solution to (\ref{15})-(\ref{16}) then $(-\theta)$ is also a solution. Hence The above analysis for (VWT) can be  extended to the case of a (VHF) flow. Consequently, multiple solutions exist if a lateral injection is applied with an injection parameter sufficiently large ($v_w=\sigma >\frac{1}{\sqrt{3}}$).\\

\noindent $\bullet$ {\bf The (VHTC) flow}\\

\noindent Consider now the case of (VHTC) flow (\ref{15})-(\ref{16})-(c). In such situation the dimensionless heat transfer coefficient is given by $h_{w}(x)=x^{-1}$.  Taking into account condition (c), we look for solution satisfying
$$\theta(0)= -\theta(0)= \kappa,$$
 where $\kappa$ is positive constant. Introducing the new variables
\[\eta\rightarrow \kappa^{2}\eta,\qquad \mbox{and}\qquad \theta\rightarrow\frac{\theta(\eta)}{\kappa}.\] 
we find the new problem
\begin{equation}\label{34}
\left\{\begin{array}{l}
\theta''+\frac{\theta^{3}}{2\kappa^{2}}+\sigma_{\kappa}\theta'=0 \qquad \mbox{with}\qquad \sigma_{\kappa}=\frac{\sigma}{\kappa^{2}},\\
\\
\theta(0)=1, \quad \theta'(0)=-\kappa^{-2},\quad \theta(\infty)=0
\end{array} \right.
\end{equation}
Using the same arguments as for the (VWT) flow, we deduce 
that there exists a minimal value $\sigma_{\kappa}^{\star}$ such that problem (\ref{34}) has positive solutions  only if   $\sigma_{\kappa}\geq {\sigma_{\kappa}}^{\star}$, more precisely $\sigma_{\kappa}^{\star}\geq \frac{1}{\kappa\sqrt{3}}$. Every solution is parametrized by $\theta'(0)\in\left[-\sqrt{\sigma_{\kappa}^{2}-\frac{1}{3\kappa^{2}}},\sigma_{\kappa}\left(\sigma_{\kappa}\sqrt{3}-1\right) \right].$
\section{Conclusion}
\setcounter{equation}{0} \setcounter{theorem}{0}
\setcounter{lemma}{0} \setcounter{remark}{0}
\setcounter{corollary}{0}
In the present paper,  we have showed how to construct exact solutions by introducing an additional logarithmic term in the usual stream-function. The relevant problem has been studied via a Crocco transformation combined with an integral equation method. The following conclusions have been made as a result of our investigation:\\
$\bullet$ There is existence of a minimal value $\sigma^{\star}$ (suction or injection) such that multiple solutions exit only for $\sigma> \sigma^{\star}$ and uniqueness holds for $\sigma=\sigma^{\star}$. \\
$\bullet$ Solutions for $\sigma=\sigma^{\star}$ exhibit an exponential decay while for $\sigma > \sigma^{\star}$ all solution are decaying algebraically.

\end{document}